\documentclass[pdflatex,sn-nature]{sn-jnl}

\usepackage{graphicx}%
\usepackage{multirow}%
\usepackage{amsmath,amssymb,amsfonts}%
\usepackage{extarrows}
\usepackage{amsthm}%
\usepackage{mathrsfs}%
\usepackage[title]{appendix}%
\usepackage{xcolor}%
\usepackage{textcomp}%
\usepackage{manyfoot}%
\usepackage{booktabs}%
\usepackage{algorithm}%
\usepackage{algorithmicx}%
\usepackage{algpseudocode}%
\usepackage{listings}%

\theoremstyle{thmstyleone}%
\theoremstyle{thmstyletwo}%

\theoremstyle{thmstylethree}%

\begin{document}

\title[Immunization on Temporal Higher-Order Networks]{Immunization on Temporal Higher-Order Networks}


\author[1,4]{\fnm{Zhihao} \sur{Han}}

\author*[1,4,6,7]{\fnm{Longzhao} \sur{Liu}}\email{longzhao@buaa.edu.cn}

\author*[1,4,6,7]{\fnm{Xin} \sur{Wang}}\email{wangxin\_1993@buaa.edu.cn}

\author[3]{\fnm{Yajing} \sur{Hao}}

\author[8]{\fnm{Hongwei} \sur{Zheng}}

\author*[1,2,4,5,6,7]{\fnm{Shaoting} \sur{Tang}}\email{tangshaoting@buaa.edu.cn}

\affil[1]{\small Institute of Artificial Intelligence, Beihang University, Beijing 100191, China}
\affil[2]{\small Hangzhou International Innovation Institute, Beihang University, Hangzhou 311115, China}
\affil[3]{\small School of Mathematical Sciences, Beihang University, Beijing, 100191, China}
\affil[4]{\small Key laboratory of Mathematics, Informatics and Behavioral Semantics, Beihang University, Beijing 100191, China}
\affil[5]{\small Institute of Medical Artificial Intelligence, Binzhou Medical University, Yantai 264003, China}
\affil[6]{\small Beijing Advanced Innovation Center for Future Blockchain and Privacy Computing, Beihang University, Beijing 100191, China}
\affil[7]{\small State Key Laboratory of Complex \& Critical Software Environment, Beihang University, Beijing 100191, China}
\affil[8]{\small Beijing Academy of Blockchain and Edge Computing, Beijing 100085, China}


\abstract{Network immunization is a powerful tool for controlling contagion processes ranging from infectious diseases to misinformation diffusion. While prior works have focused on pairwise or static networks, immunization dynamics in temporal higher-order networks remain poorly understood. Here, we introduce immunization strategies and develop a theoretical framework tailored for such temporal systems. Firstly, we reveal bistability and discontinuous transitions in prevalence as the immunization fraction varies. This implies that immunization effectiveness depends on the initial prevalence, marking a fundamental departure from pairwise networks. Building on this prevalence-dependent behavior, we propose the High Infection Contribution (HIC) strategy, demonstrating its superior performance over all evaluated heuristic strategies. Furthermore, we introduce egocentric strategies by leveraging solely local observations. Notably, the optimal egocentric strategy shifts with the contagion prevalence. Our work advances the understanding of network immunization, paving the way for effective contagion control in temporal higher-order networks.}




\maketitle

\section{Introduction}\label{introduction}

Controlling contagion processes under limited resources has become a central challenge in public health, information security and social governance. Processes such as infectious disease transmission, misinformation diffusion and panic spreading can overwhelm healthcare capacity, weaken institutional trust and disrupt social stability \cite{funk2010modelling, del2016spreading, lazer2018science, fu2021propagation, allcott2017social, azzimonti2023social}. The COVID-19 pandemic further highlighted how epidemiological and social contagion can interact across distinct but coupled structures, thereby amplifying systemic risk \cite{gallotti2020assessing, kouzy2020coronavirus}. These challenges motivate the development of network immunization strategies to mitigate contagion.

For contagion on pairwise networks, immunization has been studied extensively. Node-focused immunization, such as vaccination and isolation, protects or removes selected individuals, commonly guided by topology-based rankings such as network centrality \cite{holme2002attack, pastor2002immunization, cohen2003efficient, kitsak2010identification, liu2014controlling}. When global network information is unavailable, more deployable approaches use local observations. A prominent example is acquaintance immunization \cite{cohen2003efficient}. Edge-focused immunization aims to disrupt transmission pathways by restricting specific venues, routes or interaction channels. These strategies are typically guided by metrics that rank influential links, such as edge betweenness or epidemic importance \cite{holme2002attack, van2011decreasing, schneider2011suppressing, bishop2011link, matamalas2018effective}.

Recently, empirical evidence increasingly suggests that many natural and social systems are driven by higher-order interactions, where more than two entities interact simultaneously. The complex structure is naturally captured by hyperedges or simplices \cite{benson2016higher, grilli2017higher, mayfield2017higher, battiston2021physics, boccaletti2023structure}. In the context of contagion, nonlinear higher-order dynamical mechanisms can induce emergent behaviors, including discontinuous transitions, bistability and hysteresis, which are fundamentally absent in purely pairwise models \cite{iacopini2019simplicial, de2020social, st2021universal, st2022influential, ferraz2023multistability, ferraz2024contagion}. Moreover, contagion processes are profoundly shaped by higher-order structural properties \cite{landry2020effect, kim2023contagion, kim2024higher, burgio2024triadic, malizia2025hyperedge}. For example, increasing hyperedge overlap can shift a phase transition from discontinuous to continuous \cite{malizia2025hyperedge}. More critically, higher-order interactions are inherently temporal. Groups continuously form, dissolve, and reorganize, substantially altering epidemic thresholds and spreading outcomes \cite{chowdhary2021simplicial, han2024probabilistic, iacopini2024temporal, gallo2024higher, liu2025higher, burgio2025characteristic}. Consequently, the interplay of nonlinear group mechanisms, higher-order structures, and temporal dynamics renders  pairwise immunization strategies fundamentally inadequate.

Recent studies have begun to explore the control of higher-order contagion \cite{jhun2021effective, li2022immunization, nie2023voluntary, yang2025centrality}. Existing strategies typically prioritize nodes or groups using static centrality measures or hyperedge risk scores. However, these approaches face two critical limitations: they overlook the temporal nature of higher-order interactions and rely on global structural information which is rarely accessible in large-scale empirical systems \cite{iacopini2024temporal, gallo2024higher}. Therefore, understanding immunization dynamics in temporal environments and designing practical strategies remain open challenges.

In this work, we develop a theoretical framework for immunization on temporal higher-order networks and propose immunization strategies. First, we find the discontinuous transitions and bistability. The latter indicates that immunization effectiveness largely relies on intervention timing or initial conditions, fundamentally differing from the findings in pairwise situations. Motivated by this initial prevalence dependence, we introduce the High Infection Contribution (HIC) strategy, which outperforms all evaluated heuristic strategies. Furthermore, we develop egocentric strategies based on local observations from probes. The results demonstrate that their relative effectiveness depends heavily on the initial prevalence: Egocentric Pairwise Strategies (EPS) are more effective under low prevalence, whereas Egocentric Higher-order Strategies (EHS) excel in high-prevalence scenarios. This underscores the need to dynamically adapt immunization strategies across different phases of the contagion. Finally, we validate the robustness of our findings on real-world temporal networks. Our work unveils unique dynamical features and provides new insights into immunization strategies for temporal higher-order networks.

\section{Results}

\subsection{Theoretical framework for immunization on temporal higher-order networks}

We consider immunization for a higher-order contagion process on hypergraphs $\left(\mathcal{V}, \{\mathcal{H}(t)\}_{t=1}^T\right)$, where $\mathcal{V}$ denotes the node set and $\mathcal{H}(t)$ is a set of hyperedges at time $t$. We generate $\{\mathcal{H}(t)\}$ using the higher-order activity-driven (HOAD) model \cite{di2024percolation}  (see details in Methods). Each individual $i$ is assigned an activity vector $\mathbf{a_i}=\big(a_i^{\left(1\right)},a_i^{\left(2\right)},\ldots,a_i^{\left(M\right)}\big)$, where $a_i^{\left(m\right)}$ quantifies the propensity per unit time for individual $i$ to initiate an interaction involving $m{+}1$ individuals (i.e., a hyperedge of size $m+1$).

The contagion process has three states: susceptible (S), infected (I), or immunized (R). Infected individuals recover to susceptible with rate $\mu$. For infection, we adopt nonlinear higher-order contagion mechanism: for a hyperedge of size $m+1$, a susceptible member becomes infected with rate $\beta_m$ if the others are all infected \cite{iacopini2019simplicial}. Moreover, to account for the time delay of immunization, we introduce a time parameter $t_0$, at which point the epidemic is detected and mass vaccination is rolled out.

We then develop the theoretical framework for this process. Let $N_{\mathbf{a}}$ be the number of individuals with activity vector $\mathbf{a}$ and $n_{\mathbf{a}}=N_{\mathbf{a}}/N$ its density. Let $s_{\mathbf{a}}^{t}$, $i_{\mathbf{a}}^{t}$, and $r_{\mathbf{a}}^{t}$ denote the densities of susceptible, infected, and immunized individuals at time $t$, respectively. Thus $s_{\mathbf{a}}^{t}=n_{\mathbf{a}}-r_{\mathbf{a}}^{t}-i_{\mathbf{a}}^{t}$. We denote $\rho_0=\rho^{t_0^-}$ the prevalence immediately before immunization, where $\rho^{t}=\int i_{\mathbf{a}}^{t}\,\mathrm d\mathbf{a}$. Let $q_\mathbf{a}$ represent the immunized fraction of individuals with activity vector $\mathbf{a}$. Then, we have $r_{\mathbf{a}}^{t}=q_{\mathbf{a}}n_{\mathbf{a}}$. The overall immunization fraction satisfies $\omega=\int q_{\mathbf{a}}n_{\mathbf{a}}\,\mathrm d\mathbf{a}$. Although acquiring immunity typically involves a physiological time lag for antibody generation, for the sake of analytical tractability, we assume the immunization effect as an instantaneous event. As a result, the infection density is thinned abruptly at time $t_0$, expressed as $\rho_0^+=\rho^{t_0^+}=\int (1-q_{\mathbf{a}})\,i_{\mathbf{a}}^{t_0^-}\,\mathrm d\mathbf{a}$.

Utilizing a mean-field approximation and a discrete-time description with $\Delta t=1$, we obtain the infected activity moments
$\Theta_m^{t}=\int a^{(m)} i_{\mathbf{a}}^{t}\,\mathrm d\mathbf{a}$ for $m=1,\ldots,M$. The dynamical equation for this process can be written as
\begin{align}
i_{\mathbf{a}}^{t+1}-i_{\mathbf{a}}^{t}
=
-\mu i_{\mathbf{a}}^{t}
+ (n_{\mathbf{a}}-i_{\mathbf{a}}^{t}-r_{\mathbf{a}}^{t})
\sum_{m=1}^{M}
\beta_m
\Big[
a^{(m)}\big(\rho^{t}\big)^{m}
+
m\,\Theta_m^{t}\big(\rho^{t}\big)^{m-1}
\Big].
\label{HOAD-equation}
\end{align}
Here $a^{(m)}(\rho^t)^m$ accounts for infections when an active susceptible individual initiates a $(m{+}1)$-hyperedge and meets $m$ infected partners. $m\,\Theta_m^t(\rho^t)^{m-1}$ captures infections of susceptible individuals recruited into hyperedges initiated by infected individuals.

Without loss of generality, we focus on the simple but representative cases ($M<3$), i.e., $\mathbf{a}=\big(a^{(1)},a^{(2)}\big)$. We refer to $a^{(1)}$ and $a^{(2)}$ as the pairwise and higher-order activity rates, respectively. $\beta_1$ and $\beta_2$ represent the pairwise and higher-order infection rates. The infection intensity satisfies,
\begin{align}
\Lambda_{\mathbf a}(\rho,\Theta_1,\Theta_2)
=
\beta_1\!\left(a^{(1)}\rho+\Theta_1\right)
+
\beta_2\!\left(a^{(2)}\rho^{2}+2\Theta_2\rho\right),
\label{eq:Lambda}
\end{align}
Eq.~(\ref{HOAD-equation}) can be rewritten as
\begin{align}
i_{\mathbf a}^{t+1}-i_{\mathbf a}^{t}
=
-\mu i_{\mathbf a}^{t}
+
\big(n_{\mathbf a}-r_{\mathbf a}^{t}-i_{\mathbf a}^{t}\big)\,
\Lambda_{\mathbf a}\!\left(\rho^{t},\Theta_1^{t},\Theta_2^{t}\right).
\label{eq:HOAD-M2}
\end{align}

For a given pre-immunization prevalence $\rho_0$, we define the immunization threshold $\omega_c\left(\rho_0\right)$ as the minimum immunization fraction that prevents the system from entering an active endemic state. If the vaccination can be taken in the very early stage, i.e., $\rho_0\ll 1$, $\omega_c\left(\rho_0\right)$ reduces to the threshold $\omega_L=\lim_{\rho_0\rightarrow 0^+} \omega_c\left(\rho_0\right)$. In this limit, higher-order infection terms can be neglected. Through analyzing Eq.~(\ref{eq:HOAD-M2}) (see Methods for details), $\omega_L$ can be written as,

\begin{align}
    \sqrt{(1-\omega_L)\Psi_2}=\frac{\mu}{\beta_1}-\Psi_1
    \label{HOAD-invasion-threshold}
\end{align}
where $\Psi_x=\int (a^{(1)})^x(1-q_\mathbf{a})n_\mathbf{a}\mathrm{d}\mathbf{a}$ is the $x$-th moment of the pairwise activity rate computed over the remaining (non-immunized) population. For finite $\rho_0$, we obtain $\omega_c(\rho_0)$ numerically by solving the fixed points of Eq.~(\ref{eq:HOAD-M2}) and analyzing their stability, as detailed in the Methods.

\begin{figure}[H]
\centering
\includegraphics[width=0.9\textwidth]{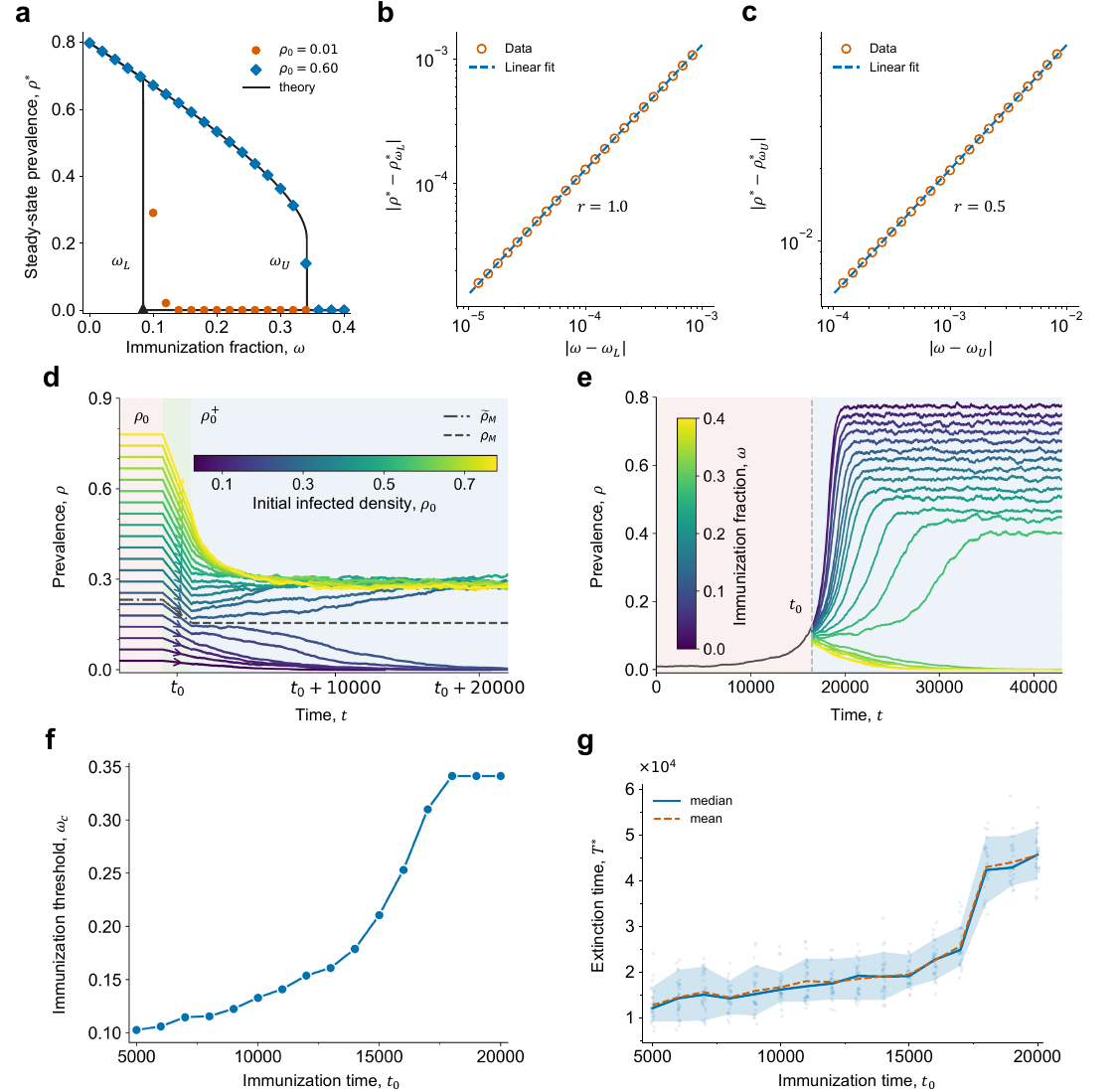}
\caption{\textbf{Immunization-induced discontinuous transitions, bistability, and the impact of immunization timing on temporal higher-order networks.}
All panels show results based on Random (R) strategy.
\textbf{a,} Steady-state prevalence $\rho^*$ is shown against immunization fraction $\omega$ at an early stage (initial infected density $\rho_0=0.01$) and a late stage ($\rho_0=0.60$). Theoretical predictions (solid curves) agree well with simulations (symbols). Notably, there emerges discontinuous transitions near the critical points. The early-immunized threshold $\omega_L$ is obtained by solving Eq.~(\ref{HOAD-invasion-threshold}), while late-immunized threshold $\omega_U$ is numerically solved.
\textbf{b,c,} Critical scaling. The panels show that $|\rho^*-\rho^*_{\omega_L}|\propto |\omega-\omega_L|$ and $|\rho^*-\rho^*_{\omega_U}|\propto |\omega-\omega_U|^{0.5}$, demonstrating that the phase transitions are hybrid.
\textbf{d,} Bistability phenomenon. For a fixed $\omega=0.33$, time evolutions are shown under different $\rho_0$ represented by different colors. Here, we assume the immunization occurs at time $t_0$. Notably, bistability occurs, where trajectories can converge to either the extinction or the endemic state.
\textbf{e,} Prevalence trajectories are shown under immunization time $t_0=16500$ indicated by the dashed line. The curves are colored by the immunization fraction $\omega$.
\textbf{f,} Immunization threshold $\omega_c$ is shown against $t_0$. The results show that the immunization threshold rises markedly as $t_0$ increases.
\textbf{g,} Extinction time $T^*$ versus $t_0$. $T^*$ is defined as the epidemic duration after the intervention. We set $\omega=0.36$. For each $t_0$, we run $30$ independent realizations. The lines show the median (blue) and mean (orange) over realizations, and the shaded band indicates the $Q_{0.10}$--$Q_{0.90}$ range. Parameters: $\beta_1=0.0034$, $\beta_2=0.06$ and $\mu=0.001$.
}
\label{figure1}
\end{figure}

\subsection{Discontinuous transitions, bistability and the effect of immunization time}

We first explore the emergent behaviors when considering immunization on temporal higher-order networks. Without loss of generality, we use Random (R) as immunization strategy. The pairwise and higher-order activity rates independently follow power-law distributions, i.e., $f_1(a^{(1)}) \propto (a^{(1)})^{-2.1}$ with $\langle a^{(1)}\rangle=0.13$ and $f_2(a^{(2)}) \propto (a^{(2)})^{-2.1}$ with $\langle a^{(2)}\rangle=0.03$.

Fig.~\ref{figure1}a shows the prevalence $\rho^*$ against immunization fraction $\omega$ under varying initial infected density $\rho_0$, where $\rho_0 = 0.01$ simulates immunization at the early stage of an outbreak, while $\rho_0=0.60$ represents the late stage. In both cases, as $\omega$ increases, the prevalence first gradually decreases, and then abruptly drops to zero once $\omega$ exceeds the immunization thresholds (denoted by $\omega_L$ and $\omega_U$ respectively). This demonstrates the emergence of discontinuous phase transition, which fundamentally differs from the findings in temporal pairwise networks \cite{liu2014controlling}. More critically, the discontinuous transition implies the drastic gap in prevalence before and after the critical point, underscoring the importance of reaching the immunization threshold. Furthermore, we explore the critical scaling around the thresholds in Fig.~\ref{figure1}b and Fig.~\ref{figure1}c. Results show $|\rho^*-\rho^*_{\omega_L}|\propto |\omega-\omega_L|$ and $|\rho^*-\rho^*_{\omega_U}|\propto |\omega-\omega_U|^{0.5}$, which indicates that the phase transitions are hybrid. In addition, we observe the bistable region in Fig.~\ref{figure1}a, where the effectiveness of immunization depends heavily on the initial infection density $\rho_0$. Specifically, there exists a threshold $\widetilde{\rho}_M$ (see Fig.~\ref{figure1}d). For $\rho_0<\widetilde{\rho}_M$, infectious disease goes extinct, whereas it becomes endemic for $\rho_0>\widetilde{\rho}_M$. Such dependence of immunization on initial conditions is another key feature absent in temporal pairwise networks. 

This dependence further enables us to explore a real-world problem: how the timing of immunization initiation influences its overall effectiveness. Fig.~\ref{figure1}e shows the prevalence trajectories, where $t_0$ represents the immunization time and each trajectory corresponds to a certain immunization fraction $\omega$. Notably, there exists an immunization threshold $\omega_c$, above which the system reaches extinction state and just below which it becomes an endemic state. Furthermore, Fig.~\ref{figure1}f shows that as $t_0$ increases, the threshold $\omega_c$ rises markedly. Moreover, even if the immunization fraction is sufficiently large to eradicate the disease, delaying the intervention, i.e., increasing $t_0$, inevitably prolongs the epidemic duration. Overall, these results demonstrate the pivotal role of early intervention in controlling epidemics.

\subsection{Targeted immunization strategies}

We explore targeted immunization strategies that assume global information is available. Motivated by activity-based targeting on temporal pairwise networks \cite{liu2014controlling}, we consider three heuristic strategies. The total activity (TA) strategy immunizes individuals in descending order of $a^{(1)}+a^{(2)}$. The higher-order activity (HA) strategy prioritizes individuals with large $a^{(2)}$, whereas the pairwise activity (PA) strategy targets those with large $a^{(1)}$.

Beyond these heuristic strategies, driven by the fundamental objective of suppressing infection, we propose the High Infection Contribution (HIC) strategy based on evaluating individual spreading capabilities (see details in Methods). Specifically, we consider the one-step increase in prevalence immediately after immunization at time $t_0$, i.e., $\Delta \rho_0^+=\rho^{t_0^++1}-\rho^{t_0^+}$.

Let $q_{\mathbf a}$ be the immunized fraction of individuals in class $\mathbf a=(a^{(1)},a^{(2)})$, and let
$n_{\mathbf a}^{\mathrm{rem}}=(1-q_{\mathbf a})n_{\mathbf a}$ denote the remaining non-immunized population. The fraction of remaining nodes is
$\chi=\int n_{\mathbf a}^{\mathrm{rem}}\,\mathrm d\mathbf a=1-\omega$. We approximate the pre-immunization infection profile as homogeneous across activity classes, i.e.,
$i_{\mathbf a}^{t_0^-}\approx \rho_0 n_{\mathbf a}$. After immunization, the infected and susceptible densities among the remaining population can be approximated as
\begin{align}
i_{\mathbf a}^{t_0^+}\approx \frac{\rho_0^+}{\chi}n_{\mathbf a}^{\mathrm{rem}},
\quad
s_{\mathbf a}^{t_0^+}\approx \frac{\chi-\rho_0^+}{\chi}n_{\mathbf a}^{\mathrm{rem}}.
\end{align}

Substituting these approximations into Eq.~(\ref{eq:HOAD-M2}) and integrating over all activity classes yields
\begin{align}
\Delta\rho_0^+
\approx
-\mu\rho_0^+
+
\frac{\chi-\rho_0^+}{\chi}
\Big[
2\beta_1\rho_0^+ B_1(q)
+
3\beta_2(\rho_0^+)^2 B_2(q)
\Big],
\end{align}
where
\begin{align}
B_m(q)=\int a^{(m)}(1-q_{\mathbf a})n_{\mathbf a}\,\mathrm d\mathbf a,
\quad m=1,2.
\end{align}

Therefore, minimizing the instantaneous infection growth is equivalent to minimizing the $q_{\mathbf a}$-dependent term
\begin{align}
J(q)=2\beta_1 B_1(q)+3\beta_2\rho_0^+B_2(q).
\end{align}

Since $B_m(q)$ measures the activity of remaining nodes after immunization, minimizing $J(q)$ is equivalent to maximizing the weighted activity removed by immunization,
\begin{align}
\int \mathrm{IC}(\mathbf a)q_{\mathbf a}n_{\mathbf a}\,\mathrm d\mathbf a,
\end{align}
where $\mathrm{IC}(\mathbf a)$ quantifies the contribution of individuals in activity class $\mathbf a$ to the instantaneous infection growth. Therefore, the IC score serves as a node importance measure for immunization, with larger values indicating greater spreading capability. The infection contribution (IC) score is given by
\begin{align}
\mathrm{IC}\!\left(a^{(1)},a^{(2)}\right)
=
2\beta_1 a^{(1)}
+
3\beta_2\rho_0(1-\omega)a^{(2)}.
\label{HOAD-IC}
\end{align}

The first term captures pairwise infection events. The second term captures higher-order infection event, which depend on the post-immunization prevalence approximated by $\rho_0^+\approx \rho_0(1-\omega)$. Accordingly, the HIC strategy prioritizes individuals with larger IC values for immunization.

\begin{figure}[h]
\centering
\includegraphics[width=0.9\textwidth]{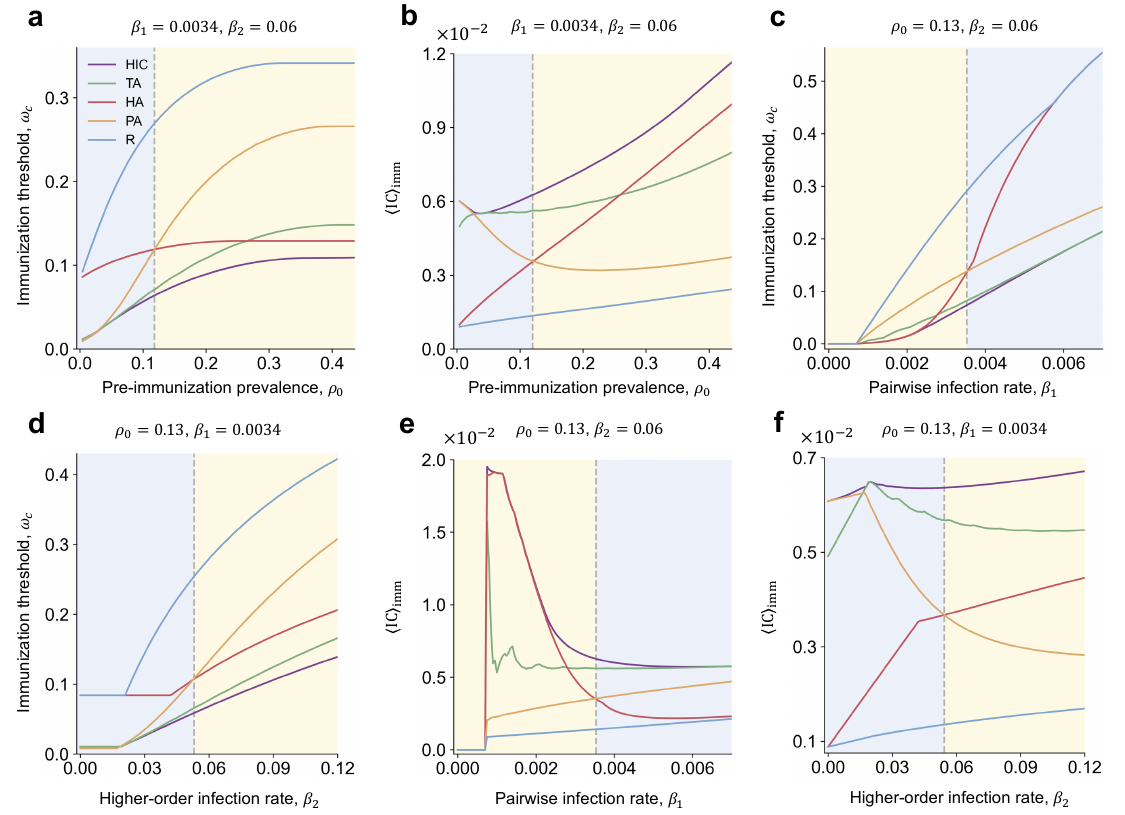}
\caption{\textbf{Effectiveness of strategies under varying initial conditions and infection rates.}
We compare five strategies (HIC, TA, HA, PA and R).
\textbf{a, c, d,} Immunization threshold $\omega_c$ is shown against initial condition $\rho_0$ (\textbf{a}), the pairwise infection rate $\beta_1$ (\textbf{c}), and the higher-order infection rate $\beta_2$ (\textbf{d}). Parameters held fixed in each sweep are indicated above the corresponding panel.
\textbf{b, e, f,} Mean infection contribution of the immunized individuals at the strategy-specific threshold, $\langle \mathrm{IC}\rangle_{\mathrm{imm}}$, is shown under the same parameters as in \textbf{a}, \textbf{c}, and \textbf{d}, respectively.
Across all cases, the HA and PA strategies exhibit a crossover. In the blue region PA achieves a smaller $\omega_c$ than HA, whereas in the yellow region HA becomes more effective than PA.}
\label{figure3}
\end{figure}

We first explore the effectiveness of targeted strategies under different initial infection density $\rho_0$ corresponding to various intervention timing. Fig.~\ref{figure3}a shows the immunization threshold $\omega_c$ required to eradicate the disease as a function of $\rho_0$. In all cases, the proposed HIC consistently yields the lowest $\omega_c$, indicating it is the optimal strategy. Moreover, the heuristic strategies (TA, HA, PA) also substantially outperform R. In particular, their relative performance depends on initial conditions: PA requires a lower $\omega_c$ than HA for small $\rho_0$, while HA outperforms PA for large $\rho_0$. This HA-PA performance crossover can be well explained by the proposed IC index, which are as follows. Around the threshold $\omega_c$, the mean IC of immunized individuals satisfies $\langle \mathrm{IC}\rangle_{\mathrm{imm}}=\frac{1}{\omega_c}\int \mathrm{IC}\left(\mathbf{a}\right)\,q_{\mathbf{a}}n_{\mathbf{a}}\,\mathrm{d}\mathbf{a}$, where $\mathrm{IC}\left(\mathbf{a}\right)$ is given by Eq.~(\ref{HOAD-IC}). Fig.~\ref{figure3}b, through comparison with Fig.~\ref{figure3}a, shows that $\langle \mathrm{IC}\rangle_{\mathrm{imm}}$ closely tracks the ranks of $\omega_c$. For low $\rho_0$, pairwise infections dominate because higher-order infection requires exposure to multiple infected individuals. Accordingly, by targeting large $a^{(1)}$, PA yields a larger $\langle \mathrm{IC}\rangle_{\mathrm{imm}}$ than HA. As $\rho_0$ grows, frequent higher-order events erode this advantage. Ultimately, at large $\rho_0$, higher-order effects dominate, making HA superior.

We further examine how the pairwise and higher-order infection rates modulate the effectiveness of these strategies (Figs.~\ref{figure3}c, d). The results show a similar picture: HIC consistently performs the best, and HA and PA still exhibit a crossover. Increasing $\beta_1$ amplifies the influence of pairwise infections, shifting the advantage to PA, whereas increasing $\beta_2$ strengthens higher-order effects, making  HA superior. The corresponding $\langle \mathrm{IC}\rangle_{\mathrm{imm}}$ curves in Figs.~\ref{figure3}e, f are also consistent with the results. Overall, the proposed HIC substantially improves immunization effectiveness, and the IC index establishes a reliable benchmark to evaluate other strategies.

\subsection{Egocentric sampling strategies}

In real-world scenarios, complete information about individuals' attributes is often unavailable, and immunization decisions must rely on observations collected from a subset of individuals over a finite time window \cite{cohen2003efficient, liu2014controlling}. This motivates the design of immunization strategies based on egocentric structural information extracted from observed interactions.

\begin{figure}[h]
\centering
\includegraphics[width=0.9\textwidth]{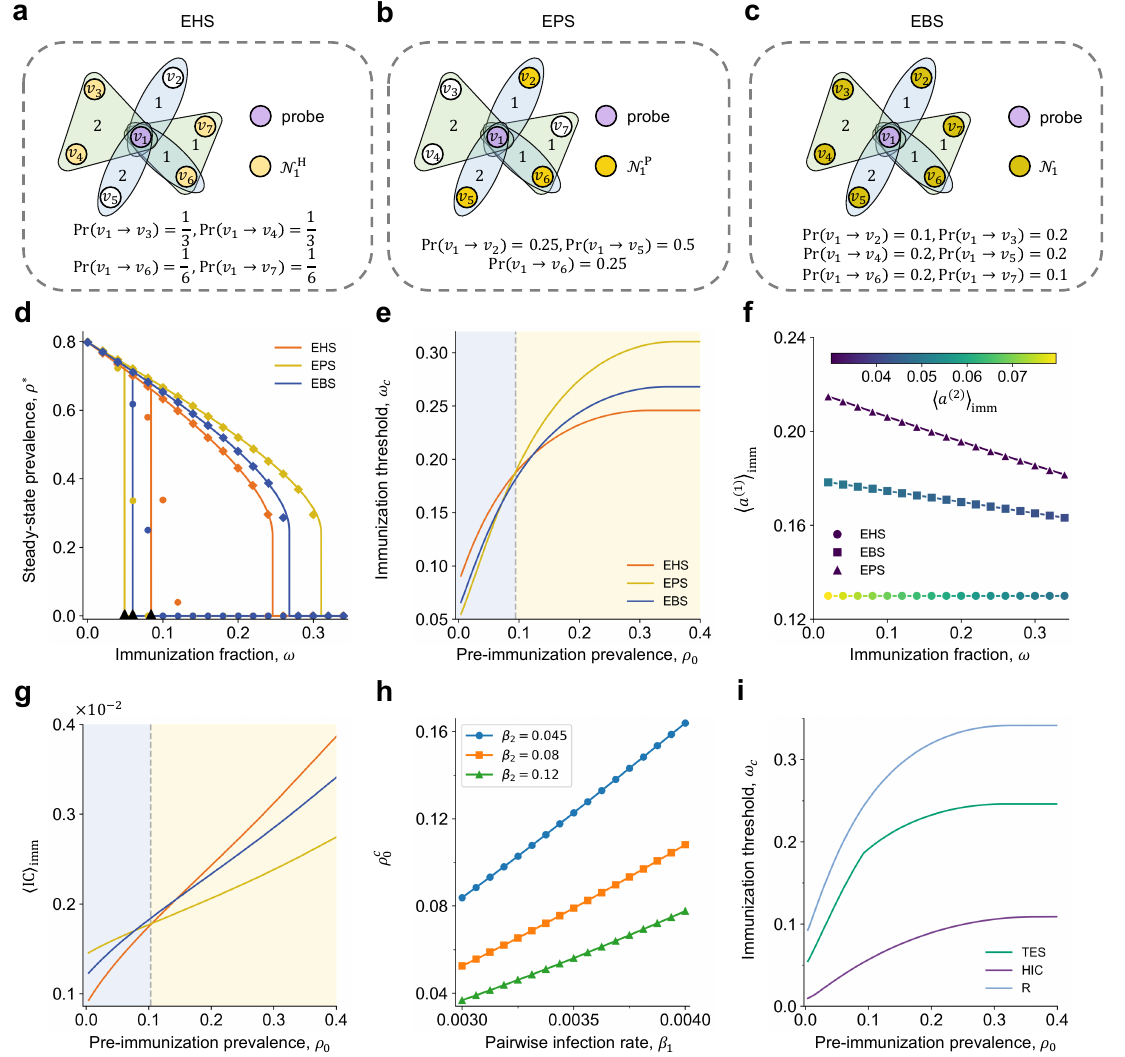}
\caption{\textbf{Egocentric sampling strategies on temporal higher-order networks.}
\textbf{a--c,} Schematic of calculating nomination probability under egocentric strategies: egocentric higher-order sampling (EHS, \textbf{a}), egocentric pairwise sampling (EPS, \textbf{b}), and egocentric balanced sampling (EBS, \textbf{c}).
\textbf{d,} Steady-state prevalence $\rho^*$ is shown against immunization fraction $\omega$ under EHS, EPS, and EBS. Symbols represent simulations and solid lines denote theoretical predictions. The results show discontinuous transitions and bistability.
\textbf{e,} Immunization threshold $\omega_c$ is shown under varying initial conditions $\rho_0$ for the three egocentric strategies. EPS is more effective at low $\rho_0$, whereas EHS becomes more effective at high $\rho_0$. EBS typically exhibits intermediate performance. The vertical dashed line marks the crossover $\rho_0^{c}$.
\textbf{f,} Mean activities of immunized individuals. The results show that EPS favors nodes with large $a^{(1)}$, whereas EHS prioritizes those with large $a^{(2)}$.
\textbf{g,} Mean infection contribution of immunized individuals $\langle \mathrm{IC}\rangle_{\mathrm{imm}}$.
\textbf{h,} The crossover prevalence $\rho_0^{c}$ is shown against $\beta_1$ under varying $\beta_2$.
\textbf{i,} Immunization threshold $\omega_c$ for the two-stage egocentric strategy (TES) compared with the HIC and R strategies.
Panels \textbf{d}, \textbf{e}, \textbf{g}, and \textbf{i} use $\beta_1=0.0034$ and $\beta_2=0.06$.}
\label{figure4}
\end{figure}

We introduce three egocentric sampling strategies that prioritize individuals based on the frequency of higher-order interactions, pairwise interactions or both. We refer to these strategies as egocentric higher-order sampling (EHS), egocentric pairwise sampling (EPS), and egocentric balanced sampling (EBS).

Specifically, we randomly select a fraction $\phi$ of individuals as probes. For each probe $u$, we construct an egocentric weighted hypergraph $G_u^{\Delta T}$ based on the interaction records within a window of length $\Delta T$. Let $\mathcal{N}_u$ be the set of individuals that interact with $u$ at least once during this window. For each $j\in\mathcal{N}_u$, we define $w_{uj}^{(1)}$ as the number of pairwise contacts between $u$ and $j$, and $w_{uj}^{(2)}$ as the number of triadic interactions $\{u,j,k\}$ (for any $k\neq u,j$). Each probe nominates exactly one neighbor for immunization, and the three strategies differ in the calculation of nomination probability. In EHS, the probe considers the frequency of triadic interactions, selecting $j$ with probability $\Pr_{\text{EHS}}\left(u\rightarrow j\right)=w_{uj}^{(2)}/\sum_{v\in\mathcal{N}_u}w_{uv}^{(2)}$ (Fig.~\ref{figure4}a). In EPS, the probe uses the information of pairwise contacts and selects $j$ with probability $\Pr_{\text{EPS}}\left(u\rightarrow j\right)=w_{uj}^{(1)}/\sum_{v\in\mathcal{N}_u}w_{uv}^{(1)}$ (Fig.~\ref{figure4}b). In EBS, the probe combines both interaction types and samples $j$ with probability  $\Pr_{\text{EBS}}\left(u\rightarrow j\right)=\left(w_{uj}^{(1)}+w_{uj}^{(2)}\right)/\sum_{v\in\mathcal{N}_u}\left(w_{uv}^{(1)}+w_{uv}^{(2)}\right)$ (Fig.~\ref{figure4}c). The immunized set $\mathcal{I}$ is the union of selected individuals after removing duplicates, and the realized immunization fraction is $\omega=|\mathcal{I}|/N$.

For large $N$, the fraction of immunized individuals in class $\mathbf{a}=(a^{(1)},a^{(2)})$ under strategy $X$ satisfies a unified approximation
\begin{align}
    q_{\mathbf{a}}^{X}\approx 1-\exp\!\left[-\phi\,\mathcal{F}_{X}(\mathbf{a})\right],
\end{align}
where $\mathcal{F}_{X}(\mathbf{a})$ is the strategy-dependent nomination intensity (see Methods for details). For the three egocentric strategies,
\begin{align}
\mathcal{F}_{\mathrm{EPS}}(\mathbf a)
&=
\int
\frac{a^{(1)}+{a^{(1)}}^\prime}{{a^{(1)}}^\prime+\langle a^{(1)}\rangle}\,
n_{\mathbf a'}\,\mathrm d\mathbf a',
\nonumber\\
\mathcal{F}_{\mathrm{EHS}}(\mathbf a)
&=
\int
\frac{a^{(2)}+{a^{(2)}}^\prime+\langle a^{(2)}\rangle}{{a^{(2)}}^\prime+2\langle a^{(2)}\rangle}\,
n_{\mathbf a'}\,\mathrm d\mathbf a',
\nonumber\\
\mathcal{F}_{\mathrm{EBS}}(\mathbf a)
&=
\int
\frac{\left({a^{(1)}}^\prime+2{a^{(2)}}^\prime\right)+\left(a^{(1)}+2a^{(2)}+2\langle a^{(2)}\rangle\right)}
{\left({a^{(1)}}^\prime+2{a^{(2)}}^\prime\right)+\left(\langle a^{(1)}\rangle+4\langle a^{(2)}\rangle\right)}\,
n_{\mathbf a'}\,\mathrm d\mathbf a'.
\end{align}

Fig.~\ref{figure4}d shows the prevalence against the immunization fraction $\omega$ under the three egocentric strategies. The theoretical predictions agree well with simulations. Additionally, the results demonstrate the emergence of discontinuous transitions and bistability, which is qualitatively similar to the findings under globally targeted strategies. Fig.~\ref{figure4}e further compares the effectiveness of the three strategies under varying initial conditions $\rho_0$. Notably, we observe a crossover between EPS and EHS at $\rho_0^c$. Specifically, EPS yields a lower immunization threshold for $\rho_0<\rho_0^{c}$, whereas EHS becomes more effective for $\rho_0>\rho_0^{c}$. EBS typically exhibits intermediate performance. 

This EPS--EHS crossover can be explained as follows. As $\rho_0$ increases, the dominant driver of infection propagation shifts from pairwise to higher-order infection events. Consistently, EPS tends to immunize nodes with higher pairwise activity rates, whereas EHS prioritizes those with greater higher-order activity rates (Fig.~\ref{figure4}f). In addition, Fig.~\ref{figure4}g shows that the crossover point $\rho_0^{c}$ can be located by the mean infection contribution of the immunized individuals.

Furthermore, Fig.~\ref{figure4}h shows that $\rho_0^{c}$ increases with the pairwise infection rate $\beta_1$ and decreases with the higher-order infection rate $\beta_2$. The results can be intuitively explained as follows. For fixed $\beta_2$, strengthening $\beta_1$ increases the relative importance of pairwise infection events, thereby inducing larger $\rho_0^c$. Similarly, for fixed $\beta_1$, strengthening $\beta_2$ enhances higher-order effects and causes smaller $\rho_0^c$.

Overall, we highlight that the effectiveness of immunization strategies is sensitive to initial conditions in temporal higher-order networks, which is fundamentally distinct from their behavior in temporal pairwise networks. Motivated by this finding, we introduce a two-stage egocentric sampling strategy (TES). Specifically, TES applies EPS for $\rho_0<\rho_0^{c}$ and transitions to EHS for $\rho_0>\rho_0^{c}$. Fig.~\ref{figure4}i presents the immunization threshold under TES. While TES is less effective than global strategy HIC, it substantially outperforms R, demonstrating that effective control remains achievable using only local structure.

\subsection{Validation on a real-world temporal higher-order network}

\begin{figure}[h]
\centering
\includegraphics[width=0.9\textwidth]{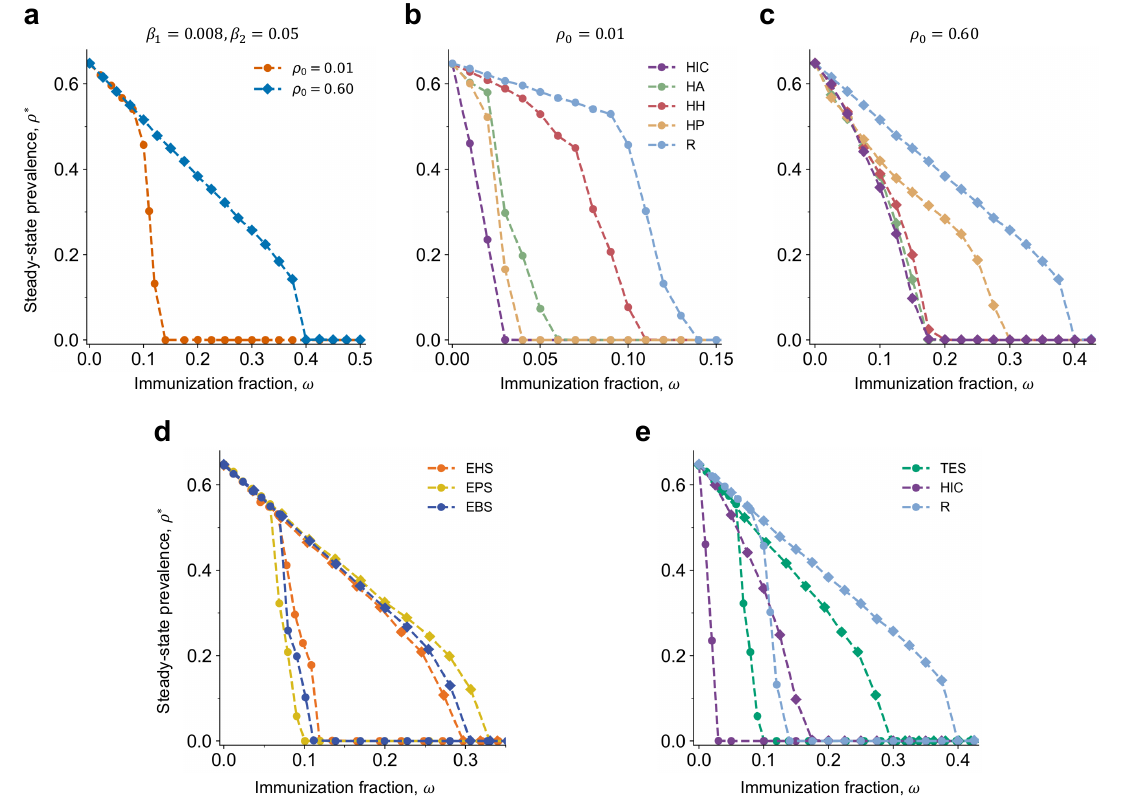}
\caption{\textbf{Validation on a real-world temporal higher-order network.}
\textbf{a,} Steady-state prevalence $\rho^*$ versus immunization fraction $\omega$ for an early stage ($\rho_0=0.01$) and a late stage ($\rho_0=0.60$) under R strategy. The results show bistability and discontinuous transitions, and a larger threshold required to guarantee extinction at the late stage.
\textbf{b,c,} Comparison of immunization strategies. Steady-state prevalence $\rho^*$ is shown against $\omega$ at $\rho_0=0.01$ (\textbf{b}) and $\rho_0=0.60$ (\textbf{c}). HIC attains the smallest immunization threshold in both stages, whereas the effectiveness of HA and PA reverses under varying $\rho_0$.
\textbf{d,} Steady-state prevalence under the egocentric sampling strategies (EHS, EPS and EBS). We find that EHS is the most effective for large $\rho_0$, while EPS performs the best for small $\rho_0$.
\textbf{e,} Comparison of the two-stage egocentric strategy (TES) with HIC and R. The results also show that stage-adaptive egocentric sampling substantially outperforms R strategy while remaining bounded by HIC. Parameters: $\beta_1=0.008$, $\beta_2=0.05$, and $\mu=0.001$.}
\label{figure5}
\end{figure}

To validate the robustness of our findings, we analyze the immunization on an empirical temporal higher-order network extracted from face-to-face interactions from the SocioPatterns dataset \cite{isella2011s}. Specifically,  based on the records in the dataset, we estimate individual pairwise and triadic activity rates (see Methods for details).

First, Fig.~\ref{figure5}a consistently illustrates the discontinuous transitions and bistability phenomenon, indicating that delaying intervention time (large $\rho_0$) substantially increases the immunization threshold. Second, Figs.~\ref{figure5}b and c demonstrate that HIC consistently achieves the smallest immunization threshold. Moreover, we find that the relative performance of HA and PA reverses when initial infection density grows. More critically, Fig.~\ref{figure5}d shows the performance of egocentric strategies. The crossover between EPS and EHS is also reproduced, while EBS remains intermediate. Finally, Fig.~\ref{figure5}e demonstrates that the two-stage strategy TES substantially improves over R, confirming its effectiveness on empirical data.

\section{Discussion}\label{Discussion}

Developing network immunization approaches to mitigate contagion is a central pursuit in diverse systems ranging from biological epidemics to misinformation diffusion. However, the design and dynamics of immunization in temporal higher-order networks remain largely unexplored. In this work, we  introduce effective immunization strategies tailored for such systems and derive an analytical framework that precisely predicts the emergent behaviors observed in large-scale simulations. 

First, in contrast to pairwise networks, we reveal the unique features of immunization in temporal higher-order networks: as the immunization fraction varies, the prevalence can exhibit discontinuous phase transitions and bistability. The former indicates the strict necessity of reaching the immunization threshold for disease eradication. The latter reflects the dependence of the immunization threshold on initial prevalence, demonstrating that delaying intervention time (which leads to higher pre-immunization prevalence) severely undermines the effectiveness of the immunization strategy. Motivated by this prevalence-dependent behavior, we propose the High Infection Contribution (HIC) strategy by quantifying the spreading influence of individuals. Extensive simulations demonstrate that HIC performs the best.

Furthermore, we develop egocentric sampling strategies which solely leverage local observations rather than global network information. Under these strategies, our main findings regarding discontinuous transitions and bistability still persist. Interestingly, the optimal egocentric strategy varies across different stages of the spreading process. Specifically, egocentric pairwise strategy performs the best at the early stage (i.e., low prevalence),  whereas egocentric higher-order strategy dominates at the late stage. This motivates us to introduce the two-stage egocentric strategy (TES), whose effectiveness is verified. Finally, we validate the robustness of our findings on empirical networks.

Our work provides practical strategies and reveals unique dynamical phenomena of immunization in temporal higher-order networks, fundamentally differing from cognitions in pairwise networks. Despite this progress, our work has several limitations that are worthy of further exploration. First, we neglect correlations between pairwise and higher-order activity rates, which may influence prioritization \cite{pozzana2017epidemic}. Second, the framework can be extended to other higher-order infection rules, including power-law infection kernels and critical mass threshold models \cite{st2021universal, ferraz2023multistability}.

\section{Methods}\label{Methods}

\subsection{Higher-Order Activity-Driven (HOAD) Model}

We employ the higher-order activity-driven (HOAD) model to generate temporal hypergraphs \cite{di2024percolation}. Each node $i$ is assigned an activity vector $\mathbf{a}_i=\big(a_i^{(1)},a_i^{(2)},\ldots,a_i^{(M)}\big)$, where the $m$-order activity rate $a_i^{(m)}$ quantifies the propensity per unit time of node $i$ to initiate an interaction involving $m{+}1$ nodes (that is, a $(m{+}1)$-hyperedge). The activity components are drawn from prescribed distributions $f_m(a^{(m)})$ for $m=1,\ldots,M$. We assume independence across interaction orders, so that the joint distribution factorizes as $f(\mathbf{a})=\prod_{m=1}^{M} f_m\!\left(a^{(m)}\right)$.

The HOAD model starts from $N$ initially disconnected nodes. At each time step $t$, node $i$ randomly selects $m$ nodes with probability $a_i^{(m)}\Delta t$, forming a $(m+1)$-hyperedge. At the next time step $t+\Delta t$, all hyperedges are erased and the process continues. The temporal hypergraph is defined by the sequence $\{\mathcal H(t)\}_{t=1}^T$ of instantaneous hypergraphs.

\subsection{Theoretical analysis for the immunization threshold at the early stage}

We derive the immunization threshold $\omega_L$ in the early stage ($\rho_0\ll 1$). We focus on $M=2$ and consider an arbitrary immunization profile $q_{\mathbf a}$ applied at time $t_0$. In this setting, $r_{\mathbf a}=q_{\mathbf a}n_{\mathbf a}$ and $\omega=\int q_{\mathbf a}n_{\mathbf a}\,\mathrm d\mathbf a$. Starting from the mean-field dynamics in Eq.~(\ref{HOAD-equation}) and retaining only the leading-order contributions in $\rho^t$, the higher-order infection terms are $O\!\big((\rho^t)^2\big)$ which are negligible. Linearizing around the disease-free fixed point yields $s_{\mathbf a}^t=n_{\mathbf a}-r_{\mathbf a}^t-i_{\mathbf a}^t\approx (1-q_{\mathbf a})n_{\mathbf a}$ and
\begin{align}
i_{\mathbf a}^{t+1}-i_{\mathbf a}^{t}
=
-\mu i_{\mathbf a}^{t}
+
\beta_1 (1-q_{\mathbf a})n_{\mathbf a}\Big(a^{(1)}\rho^t+\Theta_1^t\Big),
\label{eq:inv_linear}
\end{align}
where $\rho^t=\int i_{\mathbf a}^t\,\mathrm d\mathbf a$ and $\Theta_1^t=\int a^{(1)}i_{\mathbf a}^t\,\mathrm d\mathbf a$.

Define the moments of the pairwise activity rate over the remaining (non-immunized) population,
\begin{align}
\Psi_x\equiv \int \big(a^{(1)}\big)^x (1-q_{\mathbf a})n_{\mathbf a}\,\mathrm d\mathbf a,
\qquad x=0,1,2.
\label{eq:Psi_def}
\end{align}
Thus, $\Psi_0=1-\omega$. Integrating Eq.~(\ref{eq:inv_linear}) over $\mathbf a$ gives
\begin{align}
\rho^{t+1}-\rho^{t}
=
-\mu\rho^t+\beta_1\Psi_1\rho^t+\beta_1(1-\omega)\Theta_1^t.
\label{eq:inv_rho}
\end{align}
Then, multiplying Eq.~(\ref{eq:inv_linear}) by $a^{(1)}$ and integrating yields
\begin{align}
\Theta_1^{t+1}-\Theta_1^{t}
=
-\mu\Theta_1^t+\beta_1\Psi_2\rho^t+\beta_1\Psi_1\Theta_1^t.
\label{eq:inv_theta}
\end{align}
Through linearization, Eqs. (\ref{eq:inv_rho}) and (\ref{eq:inv_theta}) define a linear map $\mathbf x^{t+1}=\mathbf M\mathbf x^t$ for $\mathbf x^t=(\rho^t,\Theta_1^t)^\top$, with
\begin{align}
\mathbf{M}=
\left(
\begin{array}{cc}
1-\mu+\beta_1\Psi_1 & \beta_1(1-\omega) \\
\beta_1\Psi_2 & 1-\mu+\beta_1\Psi_1
\end{array}
\right).
\label{eq:inv_M}
\end{align}
The disease-free fixed point is linearly stable if and only if the leading eigenvalue satisfies $\lambda_{\max}(\mathbf M)<1$. The eigenvalues are
\begin{align}
\lambda_{\pm}
=
1-\mu+\beta_1\Psi_1\pm \beta_1\sqrt{(1-\omega)\Psi_2}.
\label{eq:inv_eigs}
\end{align}
The immunization threshold $\omega_L$ is therefore determined by $\lambda_{+}=1$, which gives Eq.~(\ref{HOAD-invasion-threshold}).

\subsection{Numerical determination of the immunization threshold for finite pre-immunization prevalence}
\label{sec:methods-finite-rho0-threshold}

We numerically calculate the immunization threshold $\omega_c(\rho_0)$. Let $\rho_0$ be the infection prevalence at the immunization time $t_0$. Let $i_{\mathbf a}^{t_0^-}$ denote the pre-immunization infection density in class $\mathbf a$, and thus we have $\rho_0=\int i_{\mathbf a}^{t_0^-}\,\mathrm d\mathbf a$. Immunization strategy is specified by $q_{\mathbf a}\in[0,1]$ which yields a time-independent immunized density $r_{\mathbf a}=q_{\mathbf a}n_{\mathbf a}$ for all $t\ge t_0^+$. The overall immunization fraction is $\omega=\int q_{\mathbf a}n_{\mathbf a}\,\mathrm d\mathbf a$. For analytical tractability, we assume immunization thins infections instantaneously,
\begin{align}
i_{\mathbf a}^{t_0^+}=(1-q_{\mathbf a})\,i_{\mathbf a}^{t_0^-},
\quad
\rho_0^{+}(\omega)=\int i_{\mathbf a}^{t_0^+}\,\mathrm d\mathbf a
=\int (1-q_{\mathbf a})\,i_{\mathbf a}^{t_0^-}\,\mathrm d\mathbf a.
\label{eq:post-thinning}
\end{align}

We focus on $M=2$. At a fixed point $i_{\mathbf a}^{t+1}=i_{\mathbf a}^{t}=i_{\mathbf a}^*$, Eq.~(\ref{eq:HOAD-M2}) yields
\begin{align}
i_{\mathbf a}^*
=
\frac{\big(n_{\mathbf a}-r_{\mathbf a}\big)\,
\Lambda_{\mathbf a}\!\left(\rho^{*},\Theta_1^{*},\Theta_2^{*}\right)}
{\mu+\Lambda_{\mathbf a}\!\left(\rho^{*},\Theta_1^{*},\Theta_2^{*}\right)}.
\label{eq:fixedpoint-explicit}
\end{align}
The macroscopic fixed points $(\rho^{*},\Theta_1^{*},\Theta_2^{*})$ are obtained by substituting Eq.~(\ref{eq:fixedpoint-explicit}) into the definitions of $(\rho,\Theta_1,\Theta_2)$, i.e.,
\begin{align}
F_0(\rho,\Theta_1,\Theta_2)
&\equiv \rho-\int i_{\mathbf a}(\rho,\Theta_1,\Theta_2)\,\mathrm d\mathbf a=0,\nonumber\\
F_1(\rho,\Theta_1,\Theta_2)
&\equiv \Theta_1-\int a^{(1)} i_{\mathbf a}(\rho,\Theta_1,\Theta_2)\,\mathrm d\mathbf a=0,\nonumber\\
F_2(\rho,\Theta_1,\Theta_2)
&\equiv \Theta_2-\int a^{(2)} i_{\mathbf a}(\rho,\Theta_1,\Theta_2)\,\mathrm d\mathbf a=0,
\label{eq:self-consistency}
\end{align}
where $i_{\mathbf a}(\rho,\Theta_1,\Theta_2)$ is given by the right-hand side of Eq.~(\ref{eq:fixedpoint-explicit}). We discretize the $(a^{(1)},a^{(2)})$ space and evaluate the integrals in Eq.~(\ref{eq:self-consistency}) as weighted sums on the discretization grid. For each prescribed $\omega$, we solve Eq.~(\ref{eq:self-consistency}) from multiple initial guesses to retrieve all distinct fixed points. For each fixed point, we construct its corresponding density profile using Eq.~(\ref{eq:fixedpoint-explicit}). We then add small perturbations to this profile and iterate the dynamics in Eq.~(\ref{eq:HOAD-M2}). A fixed point is classified as stable if perturbations decay under iteration.

For a given immunized density $r_{\mathbf a}$, Eq.~(\ref{eq:self-consistency}) admits one, two, or three fixed points in $(\rho,\Theta_1,\Theta_2)$. The disease-free fixed point $\rho_L=0$ always exists. When it is the only fixed point, it is stable and trajectories converge to $\rho_L=0$. When two fixed points exist, the disease-free fixed point is unstable and the second fixed point is stable, corresponding to endemic state $\rho_U>0$. When three fixed points exist, there are the disease-free fixed point, an intermediate unstable fixed point corresponding to prevalence $\rho_M$, and a stable endemic fixed point corresponding to prevalence $\rho_U$, where $0=\rho_L<\rho_M<\rho_U$.

We calculate $\omega_c(\rho_0)$ by combining the fixed-point structure with the post-immunization initial condition in Eq.~(\ref{eq:post-thinning}). In the three-fixed-point regime, the intermediate unstable fixed point defines the basin boundary along the one-parameter family $\rho_0^{+}(\omega)=\rho_M(\omega_c)$.

To express the threshold in terms of the pre-immunization prevalence $\rho_0$, we map the intermediate fixed point back to its pre-immunization counterpart. Specifically, for each $\omega$ in the three-fixed-point regime, we obtain $(\rho_M,\Theta_1^{M},\Theta_2^{M})$ from Eq.~(\ref{eq:self-consistency}) and construct the corresponding post-immunization intermediate fixed-point density
\begin{align}
i_{\mathbf a}^{M,+}(\omega)
=
\frac{\big(n_{\mathbf a}-r_{\mathbf a}\big)\,
\Lambda_{\mathbf a}\!\left(\rho_{M}(\omega),\Theta_1^{M}(\omega),\Theta_2^{M}(\omega)\right)}
{\mu+\Lambda_{\mathbf a}\!\left(\rho_{M}(\omega),\Theta_1^{M}(\omega),\Theta_2^{M}(\omega)\right)}.
\label{eq:iMplus}
\end{align}

The pre-immunization counterpart of the intermediate fixed point is $i_{\mathbf a}^{M,-}(\omega)=i_{\mathbf a}^{M,+}(\omega)/(1-q_{\mathbf a})$. When $q_{\mathbf{a}}=1$, we regularize the inversion by replacing $q_{\mathbf a}$ with $1-10^{-8}$ to avoid division by zero. The mapped pre-immunization intermediate prevalence is $\widetilde{\rho}_M(\omega)=\int i_{\mathbf a}^{M,-}(\omega)\,\mathrm d\mathbf{a}$. This yields an equivalent threshold for a prescribed $\rho_0$,
\begin{align}
\rho_0^{+}(\omega_c)=\rho_M(\omega_c) \quad\Longleftrightarrow\quad
\rho_0=\widetilde{\rho}_M(\omega_c).
\label{eq:finite-threshold-condition}
\end{align}
We solve Eq.~(\ref{eq:finite-threshold-condition}) for $\omega$ using a bracketed root-finding procedure within the three-fixed-point regime.

If no solution of Eq.~(\ref{eq:finite-threshold-condition}) exists, the threshold is instead determined by the loss of the endemic attractor. We scan $\omega$ and identify the smallest value at which Eq.~(\ref{eq:self-consistency}) yields no non-zero stable fixed point. This value is reported as $\omega_c(\rho_0)$.

\subsection{Derivation of the high infection contribution (HIC) strategy}
\label{sec:methods-hic}

HIC strategy is derived by minimizing the instantaneous infection growth immediately after immunization at time $t_0$. Here, we focus on $M=2$. Immunization strategy is specified by $q_{\mathbf a}\in[0,1]$, so that for $t\ge t_0^+$ the immunized density is $r_{\mathbf{a}}=q_{\mathbf a}n_{\mathbf a}$. Thus, the non-immunized population density in class $\mathbf a$ is $n_{\mathbf a}^{\mathrm{rem}}=(1-q_{\mathbf a})n_{\mathbf a}$ and the non-immunized population fraction is $\chi=\int n_{\mathbf a}^{\mathrm{rem}}\,\mathrm d\mathbf a=1-\omega$.

We quantify instantaneous growth from $t_0^+$ to $t_0^++1$, which is $\Delta\rho_0^+\equiv \rho^{t_0^++1}-\rho^{t_0^+}$. Evaluating Eq.~(\ref{eq:HOAD-M2}) at $t=t_0^+$ and integrating it over $\mathbf a$ gives
\begin{align}
\Delta\rho_0^+=-\mu\rho_0^++\int s_{\mathbf a}^{t_0^+}\,
\Lambda_{\mathbf a}\!\left(\rho_0^+,\Theta_1^{t_0^+},\Theta_2^{t_0^+}\right)
\,\mathrm d\mathbf a,
\label{eq:hic_deltarho_start}
\end{align}
where $\Theta_m^{t_0^+}=\int a^{(m)}i_{\mathbf a}^{t_0^+}\,\mathrm d\mathbf a$ and $\Lambda_{\mathbf a}$ is the infection intensity given by Eq.(2).
Eq.~(\ref{eq:hic_deltarho_start}) is exact but not closed because it depends on the post-immunization density through $(s_{\mathbf a}^{t_0^+},\Theta_1^{t_0^+},\Theta_2^{t_0^+})$.

To obtain a tractable expression for $\Delta\rho_0^+$, we approximate the pre-immunization infection profile as homogeneous across activity classes, i.e., $i_{\mathbf a}^{t_0^-}\approx \rho_0 n_{\mathbf a}$. Applying the relation in Eq.~(\ref{eq:post-thinning}) yields the post-immunization infection density
\begin{align}
i_{\mathbf a}^{t_0^+}=(1-q_{\mathbf a})\,i_{\mathbf a}^{t_0^-}
\approx \rho_0\,n_{\mathbf a}^{\mathrm{rem}},
\quad
\rho_0^+=\int i_{\mathbf a}^{t_0^+}\,\mathrm d\mathbf a
\approx \rho_0 \chi.
\label{eq:hic_post_profile}
\end{align}

Equivalently, using $\rho_0^+\approx \rho_0\chi$, the post-immunization densities can be written as
\begin{align}
i_{\mathbf a}^{t_0^+}\approx \frac{\rho_0^+}{\chi}\,n_{\mathbf a}^{\mathrm{rem}},
\quad
s_{\mathbf a}^{t_0^+}=n_{\mathbf a}^{\mathrm{rem}}-i_{\mathbf a}^{t_0^+}\approx \frac{\chi-\rho_0^+}{\chi}\,n_{\mathbf a}^{\mathrm{rem}}.
\label{eq:hic_closure_t0}
\end{align}

Under Eq.~(\ref{eq:hic_closure_t0}), the infected activity moments at $t_0^+$ satisfy
$\Theta_m^{t_0^+}\approx (\rho_0^+/\chi)B_m(q)$, where
\begin{align}
B_m(q)\equiv \int a^{(m)}n_{\mathbf a}^{\mathrm{rem}}\,\mathrm d\mathbf a
=\int a^{(m)}(1-q_{\mathbf a})n_{\mathbf a}\,\mathrm d\mathbf a,\quad m=1,2.
\label{eq:hic_Bm}
\end{align}

Substituting Eqs.~(\ref{eq:hic_closure_t0})--(\ref{eq:hic_Bm}) into Eq.~(\ref{eq:hic_deltarho_start}) and using the explicit form of $\Lambda_{\mathbf a}$ in Eq.~(\ref{eq:Lambda}), we have
\begin{align}
\Delta\rho_0^+
\approx
-\mu\rho_0^++
\frac{\chi-\rho_0^+}{\chi}
\Big[
2\beta_1\rho_0^+\,B_1(q)
+
3\beta_2(\rho_0^+)^{2}\,B_2(q)
\Big].
\label{eq:hic_deltarho_closed}
\end{align}

For a prescribed $(\rho_0,\omega)$, we have $\rho_0^+=\rho_0(1-\omega)$ and $\chi=1-\omega$. Minimizing the instantaneous growth in Eq.~(\ref{eq:hic_deltarho_closed}) over the allocation $q_{\mathbf a}$ is therefore equivalent to minimizing the $q_{\mathbf a}$-dependent term
\begin{align}
J(q)
=
2\beta_1 B_1(q)
+
3\beta_2\rho_0^+ B_2(q).
\label{eq:hic_objective}
\end{align}

By expressing $B_m(q)$ as $B_m(q)=\langle a^{(m)}\rangle-C_m(q)$ with
$C_m(q)=\int a^{(m)}q_{\mathbf a}n_{\mathbf a}\,\mathrm d\mathbf a$ and
$\langle a^{(m)}\rangle=\int a^{(m)}n_{\mathbf a}\,\mathrm d\mathbf a$, the problem of minimizing $J(q)$ subject to the constraint $\omega=\int q_{\mathbf a}n_{\mathbf a}\,\mathrm d\mathbf a$ is equivalent to
\begin{align}
\max_{q_{\mathbf a}}
\int
\mathrm{IC}(\mathbf a)\,q_{\mathbf a}\,n_{\mathbf a}\,\mathrm d\mathbf a,
\quad
\mathrm{IC}(\mathbf a)
=
2\beta_1 a^{(1)}
+
3\beta_2\rho_0^{+}a^{(2)}.
\label{eq:hic_IC_derivation}
\end{align}

Therefore, minimizing $J(q)$ is equivalent to maximizing the total infection contribution removed by immunization. The HIC strategy thus ranks individuals by $\mathrm{IC}(\mathbf a)$ and immunizes those with the largest values until the prescribed immunization fraction $\omega$ is reached. Using $\rho_0^+=\rho_0(1-\omega)$ from Eq.~(\ref{eq:hic_post_profile}) yields the infection contribution ($\mathrm{IC}$) score $\mathrm{IC}(\mathbf a)=2\beta_1 a^{(1)}+3\beta_2\rho_0(1-\omega)a^{(2)}$.

\subsection{Condition for equal immunization thresholds between strategies}
\label{sec:methods-equal-threshold}

We derive a compact condition under which two strategies yield the same immunization threshold for a prescribed pre-immunization prevalence $\rho_0$. Consider strategies $X$ and $Y$, specified by immunization profiles $q_{\mathbf a}^{X}$ and $q_{\mathbf a}^{Y}$. In the three-fixed-point regime, the threshold is characterized by the condition $\rho_0^{+}(\omega_c)=\rho_M(\omega_c)$ (Sec.~\ref{sec:methods-finite-rho0-threshold}), where $\rho_M$ is the prevalence of the intermediate unstable fixed point.

At any fixed point of Eq.~(\ref{eq:HOAD-M2}), integrating over $\mathbf a$ yields the exact prevalence balance
\begin{align}
0
=
-\mu\rho^{*}
+
\int s_{\mathbf a}^{*}\,
\Lambda_{\mathbf a}\!\left(\rho^{*},\Theta_1^{*},\Theta_2^{*}\right)
\,\mathrm d\mathbf a,
\label{eq:eqth_exact_balance}
\end{align}
with $\rho^{*}=\int i_{\mathbf a}^{*}\,\mathrm d\mathbf a$, $\Theta_m^{*}=\int a^{(m)}i_{\mathbf a}^{*}\,\mathrm d\mathbf a$, and $s_{\mathbf a}^{*}=n_{\mathbf a}-r_{\mathbf a}-i_{\mathbf a}^{*}$.
Evaluating Eq.~(\ref{eq:eqth_exact_balance}) at the intermediate unstable fixed point at threshold and adopting the same homogeneous approximation and closure as in Sec.~\ref{sec:methods-hic} (Eqs.~(\ref{eq:hic_post_profile})--(\ref{eq:hic_closure_t0})), we obtain for any strategy $K$ the closed relation
\begin{align}
0
\approx
-\mu\rho_0^{+}(\omega_c^{K})
+
\frac{\chi-\rho_0^{+}(\omega_c^{K})}{\chi}
\Big[
2\beta_1\rho_0^{+}(\omega_c^{K})\,B_1^{K}(\omega_c^{K})
+
3\beta_2\big(\rho_0^{+}(\omega_c^{K})\big)^{2} B_2^{K}(\omega_c^{K})
\Big],
\label{eq:eqth_closed_single}
\end{align}
where $\chi=1-\omega_c^{K}$ and
$B_m^{K}(\omega)=\int a^{(m)}(1-q_{\mathbf a}^{K})n_{\mathbf a}\,\mathrm d\mathbf a$ for $m=1,2$.

Suppose the two strategies have equal thresholds at $\rho_0$,
$\omega_c^{X}(\rho_0)=\omega_c^{Y}(\rho_0)\equiv \omega^{*}$.
Then $\chi=1-\omega^{*}$ is shared and, under Eq.~(\ref{eq:hic_post_profile}), the post-immunization prevalence is also shared, $\rho_0^{+}(\omega^{*})\approx \rho_0(1-\omega^{*})$.
Subtracting Eq.~(\ref{eq:eqth_closed_single}) for $X$ and $Y$ evaluated at $(\chi,\rho_0^{+})$ yields
\begin{align}
2\beta_1\,\Delta B_1(\omega^{*})
+
3\beta_2\,\rho_0^{+}(\omega^{*})\,\Delta B_2(\omega^{*})
=
0,
\label{eq:eqth_condition_B_compact}
\end{align}
where $\Delta B_m(\omega^{*})\equiv B_m^{X}(\omega^{*})-B_m^{Y}(\omega^{*})$.
Using $B_m^{X}(\omega)=\langle a^{(m)}\rangle-C_m^{X}(\omega)$ with
$C_m^{X}(\omega)=\int a^{(m)}q_{\mathbf a}^{X}n_{\mathbf a}\,\mathrm d\mathbf a$,
Eq.~(\ref{eq:eqth_condition_B_compact}) is equivalently written as
\begin{align}
2\beta_1\,\Delta C_1(\omega^{*})
+
3\beta_2\,\rho_0^{+}(\omega^{*})\,\Delta C_2(\omega^{*})
=
0.
\label{eq:eqth_condition_C_compact}
\end{align}

Using the infection contribution score $\mathrm{IC}(\mathbf a)$ introduced in Sec.~\ref{sec:methods-hic} (Eq.~(\ref{eq:hic_IC_derivation})), Eq.~(\ref{eq:eqth_condition_C_compact}) can be written as
\begin{align}
\int \mathrm{IC}(\mathbf a)\,q_{\mathbf a}^{X}n_{\mathbf a}\,\mathrm d\mathbf a
=
\int \mathrm{IC}(\mathbf a)\,q_{\mathbf a}^{Y}n_{\mathbf a}\,\mathrm d\mathbf a.
\label{eq:eqth_IC_equal_compact}
\end{align}

Since both strategies immunize the same fraction $\omega^{*}$, we define the mean infection contribution among immunized individuals as
\begin{align}
\langle \mathrm{IC}\rangle_{\mathrm{imm}}^{X}(\omega^{*})
\equiv
\frac{1}{\omega^{*}}
\int \mathrm{IC}(\mathbf a)\,q_{\mathbf a}^{X}n_{\mathbf a}\,\mathrm d\mathbf a,
\quad
\langle \mathrm{IC}\rangle_{\mathrm{imm}}^{Y}(\omega^{*})
\equiv
\frac{1}{\omega^{*}}
\int \mathrm{IC}(\mathbf a)\,q_{\mathbf a}^{Y}n_{\mathbf a}\,\mathrm d\mathbf a,
\label{eq:eqth_meanIC_def}
\end{align}
so that the equal-threshold condition becomes
\begin{align}
\langle \mathrm{IC}\rangle_{\mathrm{imm}}^{X}(\omega^{*})
=
\langle \mathrm{IC}\rangle_{\mathrm{imm}}^{Y}(\omega^{*}).
\label{eq:eqth_meanIC_equal_final}
\end{align}

\subsection{Derivation of egocentric immunization profiles}
\label{sec:methods-egocentric-largeN}

We derive the immunization profiles $q_{\mathbf a}$ for three egocentric strategies. The derivation is carried out at the activity-class level $\mathbf a=(a^{(1)},a^{(2)})$ with class density $n_{\mathbf a}$. Let $N_p=\phi N$ be the number of probes. For strategy $X$, denote by $p_{\mathbf a}^{X}$ the probability that a uniformly random probe nominates an individual in class $\mathbf{a}$. Assuming independence across probes, the probability that this individual is never nominated is $(1-p_{\mathbf a}^{X})^{N_p}$, hence
\begin{equation}
q_{\mathbf a}^{X}
=
1-(1-p_{\mathbf a}^{X})^{N_p}
\approx
1-\exp\!\left(-N_pp_{\mathbf a}^{X}\right),
\label{eq:qa_from_pa_revise}
\end{equation}
where the exponential form uses $p_{\mathbf a}^{X}=O(1/N)$, shown below.

To compute $p_{\mathbf a}^{X}$, we evaluate the nomination probability from a probe in class $\mathbf a'$ to a target in class $\mathbf a$ over an observation window of length $\Delta T$. Let $\mathbb{E}[w^{(1)}(\mathbf a',\mathbf a)]$ and $\mathbb{E}[w^{(2)}(\mathbf a',\mathbf a)]$ be the expected pairwise and triadic co-occurrence counts between the two classes within the window. The expected per-step number of pairwise interactions containing both individuals is
\begin{equation}
c^{P}(\mathbf a',\mathbf a)=\frac{{a^{(1)}}'+a^{(1)}}{N-1},
\quad
\mathbb{E}\!\left[w^{(1)}(\mathbf a',\mathbf a)\right]=\Delta T\,c^{P}(\mathbf a',\mathbf a).
\label{eq:wp_expect}
\end{equation}
The expected per-step number of triadic interactions containing both individuals is
\begin{equation}
c^{H}(\mathbf a',\mathbf a)
=
\frac{2\big({a^{(2)}}'+a^{(2)}\big)}{N-1}
+
\frac{2\big(N\langle a^{(2)}\rangle-{a^{(2)}}'-a^{(2)}\big)}{(N-1)(N-2)},
\quad
\mathbb{E}\!\left[w^{(2)}(\mathbf a',\mathbf a)\right]=\Delta T\,c^{H}(\mathbf a',\mathbf a),
\label{eq:wh_expect}
\end{equation}
where $\langle a^{(m)}\rangle=\int a^{(m)} n_{\mathbf a}\,\mathrm d\mathbf a$.

For a probe in class $\mathbf a'$, the nomination probability to class $\mathbf a$ is approximated by the corresponding ratio of expected counts. In EPS, this gives
\begin{equation}
\Pr_{\mathrm{EPS}}(\mathbf a'\!\rightarrow\!\mathbf a)
\approx
\frac{\mathbb{E}[w^{(1)}(\mathbf a',\mathbf a)]}{\sum_{\mathbf b\neq \mathbf a'} \mathbb{E}[w^{(1)}(\mathbf a',\mathbf b)]}
=
\frac{\Delta T\,c^{P}(\mathbf a',\mathbf a)}{\sum_{\mathbf b\neq \mathbf a'} \Delta T\,c^{P}(\mathbf a',\mathbf b)}
=
\frac{c^{P}(\mathbf a',\mathbf a)}{\sum_{\mathbf b\neq \mathbf a'} c^{P}(\mathbf a',\mathbf b)}.
\label{eq:eps_ratio_cancel}
\end{equation}
Using $\sum_{\mathbf b\neq \mathbf a'} c^{P}(\mathbf a',\mathbf b)=\big((N-2){a^{(1)}}'+N\langle a^{(1)}\rangle\big)/(N-1)$, we obtain
\begin{equation}
\Pr_{\mathrm{EPS}}(\mathbf a'\!\rightarrow\!\mathbf a)
\approx
\frac{{a^{(1)}}'+a^{(1)}}{(N-2){a^{(1)}}'+N\langle a^{(1)}\rangle}
=
\frac{1}{N}\,
\frac{{a^{(1)}}'+a^{(1)}}{{a^{(1)}}'+\langle a^{(1)}\rangle}
+o\!\left(\frac{1}{N}\right).
\label{eq:eps_largeN_kernel}
\end{equation}

In EHS, the same cancellation of $\Delta T$ yields
\begin{equation}
\Pr_{\mathrm{EHS}}(\mathbf a'\!\rightarrow\!\mathbf a)
\approx
\frac{\mathbb{E}[w^{(2)}(\mathbf a',\mathbf a)]}{\sum_{\mathbf b\neq \mathbf a'} \mathbb{E}[w^{(2)}(\mathbf a',\mathbf b)]}
=
\frac{c^{H}(\mathbf a',\mathbf a)}{\sum_{\mathbf b\neq \mathbf a'} c^{H}(\mathbf a',\mathbf b)}.
\label{eq:ehs_ratio_cancel}
\end{equation}
Evaluating the denominator gives $\sum_{\mathbf b\neq \mathbf a'} c^{H}(\mathbf a',\mathbf b)=2\big((N-3){a^{(2)}}'+2N\langle a^{(2)}\rangle\big)/(N-1)$, hence
\begin{equation}
\Pr_{\mathrm{EHS}}(\mathbf a'\!\rightarrow\!\mathbf a)
\approx
\frac{1}{N}\,
\frac{{a^{(2)}}'+a^{(2)}+\langle a^{(2)}\rangle}{{a^{(2)}}'+2\langle a^{(2)}\rangle}
+o\!\left(\frac{1}{N}\right).
\label{eq:ehs_largeN_kernel}
\end{equation}

In EBS, the sampling weight is the sum of pairwise and triadic counts. Hence,
$\mathbb{E}[w^{(1)}(\mathbf a',\mathbf a)+w^{(2)}(\mathbf a',\mathbf a)]=\Delta T\big(c^{P}(\mathbf a',\mathbf a)+c^{H}(\mathbf a',\mathbf a)\big)$.
Therefore,
\begin{equation}
\Pr_{\mathrm{EBS}}(\mathbf a'\!\rightarrow\!\mathbf a)
\approx
\frac{c^{P}(\mathbf a',\mathbf a)+c^{H}(\mathbf a',\mathbf a)}
{\sum_{\mathbf b\neq \mathbf a'} \big(c^{P}(\mathbf a',\mathbf b)+c^{H}(\mathbf a',\mathbf b)\big)}.
\label{eq:ebs_ratio_cancel}
\end{equation}
Using the closed forms
$\sum_{\mathbf b\neq \mathbf a'} c^{P}(\mathbf a',\mathbf b)=\big((N-2){a^{(1)}}'+N\langle a^{(1)}\rangle\big)/(N-1)$ and
$\sum_{\mathbf b\neq \mathbf a'} c^{H}(\mathbf a',\mathbf b)=2\big((N-3){a^{(2)}}'+2N\langle a^{(2)}\rangle\big)/(N-1)$,
we obtain
\begin{equation}
\Pr_{\mathrm{EBS}}(\mathbf a'\!\rightarrow\!\mathbf a)
\approx
\frac{1}{N}\,
\frac{\left({a^{(1)}}'+2{a^{(2)}}'\right)+\left(a^{(1)}+2a^{(2)}+2\langle a^{(2)}\rangle\right)}
{\left({a^{(1)}}'+2{a^{(2)}}'\right)+\left(\langle a^{(1)}\rangle+4\langle a^{(2)}\rangle\right)}
+o\!\left(\frac{1}{N}\right).
\label{eq:ebs_largeN_kernel}
\end{equation}

Because probes are selected uniformly at random, the probe class $\mathbf a'$ is distributed as $n_{\mathbf a'}$. The per-probe nomination probability for class $\mathbf a$ is therefore
\begin{equation}
p_{\mathbf a}^{X}=\int \Pr_{X}(\mathbf a'\!\rightarrow\!\mathbf a)\,n_{\mathbf a'}\,\mathrm d\mathbf a'\approx
\frac{1}{N}\int f_{X}(\mathbf a',\mathbf a)\,n_{\mathbf a'}\,\mathrm d\mathbf a',
\label{eq:pa_fx}
\end{equation}
where $f_{X}$ is read off from Eqs.~(\ref{eq:eps_largeN_kernel})--(\ref{eq:ebs_largeN_kernel}). Substituting Eq.~(\ref{eq:pa_fx}) into Eq.~(\ref{eq:qa_from_pa_revise}) and using $N_p=\phi N$ yields
\begin{equation}
q_{\mathbf a}^{X}
\approx
1-\exp\!\left[-\phi\,\mathcal{F}_{X}(\mathbf a)\right],
\quad
\mathcal{F}_{X}(\mathbf a)=\int f_{X}(\mathbf a',\mathbf a)\,n_{\mathbf a'}\,\mathrm d\mathbf a'.
\label{eq:qa_unified_final}
\end{equation}
Explicitly,
\begin{align}
\mathcal{F}_{\mathrm{EPS}}(\mathbf a)
&=
\int
\frac{a^{(1)}+{a^{(1)}}^\prime}{{a^{(1)}}^\prime+\langle a^{(1)}\rangle}\,
n_{\mathbf a'}\,\mathrm d\mathbf a',
\nonumber\\
\mathcal{F}_{\mathrm{EHS}}(\mathbf a)
&=
\int
\frac{a^{(2)}+{a^{(2)}}^\prime+\langle a^{(2)}\rangle}{{a^{(2)}}^\prime+2\langle a^{(2)}\rangle}\,
n_{\mathbf a'}\,\mathrm d\mathbf a',
\nonumber\\
\mathcal{F}_{\mathrm{EBS}}(\mathbf a)
&=
\int
\frac{\left({a^{(1)}}^\prime+2{a^{(2)}}^\prime\right)+\left(a^{(1)}+2a^{(2)}+2\langle a^{(2)}\rangle\right)}
{\left({a^{(1)}}^\prime+2{a^{(2)}}^\prime\right)+\left(\langle a^{(1)}\rangle+4\langle a^{(2)}\rangle\right)}\,
n_{\mathbf a'}\,\mathrm d\mathbf a'.
\label{eq:FX_na_form}
\end{align}

\subsection{Empirical temporal higher-order network construction}
\label{sec:methods-empirical}

We use the SocioPatterns Infectious exhibition dataset of face-to-face proximity contacts, recorded at a temporal resolution of 20\,s \cite{isella2011s}. We aggregate the contact sequence into snapshots of duration $\Delta t=3$\,min. For each snapshot $t$, we construct an undirected contact graph by placing an edge $(i,j)$ if at least one contact between $i$ and $j$ occurs within the corresponding 3-min window. We then extract all 3-cliques in the snapshot graph and treat each clique $\{i,j,k\}$ as a 3-hyperedge. Together, the pairwise contacts and triadic interactions define a temporal higher-order network represented by the sequence $\{\mathcal{H}(t)\}_{t=1}^{T}$, where $T=116$. To reduce finite-size fluctuations while preserving the temporal pattern of interactions, we generate an augmented temporal hypergraph by expanding the network size at each snapshot by a factor of 100, resulting in $N=15500$ nodes \cite{young2017construction}.

We estimate each individual’s pairwise and triadic activity rates from its participation frequencies. Let $k_i^{(1)}(t)$ denote the number of pairwise contacts incident to node $i$ at snapshot $t$, and let $k_i^{(2)}(t)$ denote the number of 3-hyperedges containing node $i$ at snapshot $t$. We compute the time-averaged counts $k_i^{(m)}=T^{-1}\sum_{t=1}^{T} k_i^{(m)}(t)$ for $m\in\{1,2\}$, and use them as empirical proxies of the activity components $a_i^{(m)}$ up to a global rescaling, which does not affect rank-based immunization.

To implement the HIC strategy on empirical data, we rank nodes by their infection contribution score at $\rho_0$, $\textrm{IC}_i=\beta_1 k_i^{(1)}+\beta_2 k_i^{(2)}\,\rho_0(1-\omega)$.

\bibliography{sn-bibliography}

\end{document}